\documentclass[11pt,a4paper]{article}
\usepackage[T1]{fontenc}
\usepackage{lmodern,microtype}
\usepackage{amsmath,amssymb,bm}
\usepackage{graphicx,booktabs,placeins}
\usepackage[margin=2.25cm]{geometry}
\usepackage[font=small,labelfont=bf]{caption}
\usepackage[hidelinks]{hyperref}

\newcommand{\dd}{\mathrm{d}}
\newcommand{\GeV}{\mathrm{GeV}}
\newcommand{\tr}{\operatorname{Tr}}
\newcommand{\re}{\operatorname{Re}}
\newcommand{\as}{\alpha_s}
\newcommand{\Tmat}{\mathsf{T}}
\newcommand{\Rmat}{\mathsf{W}}
\hypersetup{pdftitle={One-loop anomalous dimensions and two-photon exchange in elastic electron-deuteron scattering},pdfauthor={Yaroslav D. Krivenko-Emetov and Iryna Myroshnykova}}
\title{One-loop anomalous dimensions and two-photon exchange in elastic electron--deuteron scattering}
\author{Yaroslav D. Krivenko-Emetov$^{1,2,*}$ and Iryna Myroshnykova$^{1}$\\[0.65em]
\parbox{0.97\textwidth}{\centering\small
$^{1}$National Technical University of Ukraine ``Igor Sikorsky Kyiv Polytechnic Institute'', Kyiv 03056, Ukraine\\
$^{2}$Institute for Nuclear Research, National Academy of Sciences of Ukraine, Kyiv 03680, Ukraine\\
$^{*}$\href{mailto:y.kryvenko-emetov@kpi.ua}{y.kryvenko-emetov@kpi.ua}}}
\date{}

\begin{document}
\maketitle
\begin{abstract}
We calculate the one-loop anomalous dimensions of the lowest local isoscalar six-quark operators and
use them in a joint analysis of cross sections and recoil polarization in elastic electron--deuteron
scattering. The lowest eigenvalues are \(1\) for helicity zero and \(2/3\) for unit helicity in the
\(\alpha_s/(2\pi)\) convention. We also use the values \(1\) and \(7/6\) from the Young-projected
calculation of Ji and Brodsky. In the one-photon approximation we express the deuteron form factors
in terms of three light-front helicity amplitudes. We add two-photon exchange to the scattering
amplitude and retain its interference with one-photon exchange in the differential cross section and
in \(t_{20}\), \(t_{21}\) and \(t_{22}\). We fit 48 cross sections and 12 tensor measurements with
\(Q^2\) between \(1\) and \(6\,\GeV^2\). The fit is most sensitive to the finite-\(Q^2\) dependence
of the three Born amplitudes. With independent finite-\(Q^2\) terms in the three amplitudes, the
one-photon fit gives \(\chi^2/\nu=1.826\). A common relative two-photon factor gives \(1.790\).
Fitting separate electric and magnetic strengths and powers reduces \(\chi^2\) by only \(0.49\) for
two additional parameters.
\end{abstract}


\section{Introduction}

Elastic electron--deuteron scattering at a few \(\GeV^2\) probes distances where both the nucleon
structure of the deuteron and the large-\(Q^2\) behavior of its electromagnetic current are
relevant~\cite{GilmanGross,BrodskyFarrar,CarlsonGross,BrodskyJi}. In the one-photon approximation
the current is described by the charge, magnetic and quadrupole form factors \(G_C\), \(G_M\) and
\(G_Q\). An unpolarized Rosenbluth separation gives two combinations of their squares,
\(A_{1\gamma}(Q^2)\) and \(B_{1\gamma}(Q^2)\). Recoil tensor polarization supplies additional
information because it depends on products of different form
factors~\cite{AbbottTensor,KobushkinPQCD}.

We consider two effects in the same analysis. The first is the logarithmic evolution of the external
distribution amplitudes in perturbative QCD. We calculate the lowest local six-quark spectrum and
compare it with the result of Ji and Brodsky. The second is the interference of one- and two-photon
exchange (TPE). Earlier analyses used phenomenological logarithmic corrections to \(A\) and
\(B\)~\cite{Shevchuk,Previous}. Here we use related functions in the Breit-frame scattering
amplitude, where the correction to the cross section and those to \(t_{20}\), \(t_{21}\) and
\(t_{22}\) follow from the same spin matrix~\cite{Kobushkin2010,Kobushkin2011,Gakh}.

We fit the cross sections at their measured angles together with the recoil tensor data. This lets
us test whether the change in the logarithmic powers can be resolved once the finite-\(Q^2\)
dependence of the one-photon helicity amplitudes and the two-photon terms are fitted to the same
measurements.

\section{Cross section and recoil tensor polarization}
\label{sec:observables}

Let \(q\) be the four-momentum transfer, \(Q^2=-q^2>0\), \(M_d\) the deuteron mass and \(\theta_{\rm
lab}\) the scattering angle of the electron in the laboratory. Put \(\eta=Q^2/(4M_d^2)\). In the
Born approximation the two structure functions are
\begin{align}
 A_{1\gamma}(Q^2)&=G_C^2+\frac89\eta^2G_Q^2+\frac23\eta G_M^2,\notag\\
 B_{1\gamma}(Q^2)&=\frac43\eta(1+\eta)G_M^2 .
 \label{eq:ab}
\end{align}
For \(\theta_{\rm lab}<180^\circ\) the reduced cross section is
\begin{equation}
 S_{1\gamma}(Q^2,\theta_{\rm lab})=
 \frac{1}{\sigma_{\rm Mott}}\left(\frac{\dd\sigma}{\dd\Omega}\right)_{1\gamma}
 =A_{1\gamma}(Q^2)+B_{1\gamma}(Q^2)\tan^2\frac{\theta_{\rm lab}}{2},
 \label{eq:reduced}
\end{equation}
where \(\sigma_{\rm Mott}\) includes recoil. The factor \(\tan^2(\theta_{\rm lab}/2)\) enhances the
sensitivity to the magnetic form factor at backward angles. The measured recoil moments are ratios
of spin-weighted to unpolarized rates. For example, the Born numerator of \(t_{20}\) contains
\(G_CG_Q\) and that of \(t_{21}\) contains \(G_MG_Q\). These products let the measurements
distinguish relative signs that are lost when the form factors are squared in Eq.~\eqref{eq:ab}. We
use the recoil convention of Ref.~\cite{AbbottTensor}. Appendix~\ref{app:spin} gives the dimensional
cross section also at \(180^\circ\), where \(\sigma_{\rm Mott}\) vanishes.

For the interference calculation it is useful to absorb the kinematic weights into the form factors,
\begin{equation}
 \bm v=(v_C,v_Q,v_M)
 =\bigl(G_C,\sqrt{8/9}\,\eta G_Q,\sqrt{2\eta/3}\,G_M\bigr).
 \label{eq:v}
\end{equation}
We write the electric and magnetic parts of \(A_{1\gamma}\) as \(A_E=v_C^2+v_Q^2\) and
\(A_M=v_M^2\). In this notation \(B_{1\gamma}=2(1+\eta)A_M\).

\section{One-photon helicity amplitudes}
\label{sec:born}

Work in a light-front frame with \(q^+=0\) and \(p^+=p'^+=P^+\), where \(p\) and \(p'\) are the
initial and final deuteron momenta and \(p^+=p^0+p^3\). We use
\(H_{\lambda'\lambda}=J^+_{\lambda'\lambda}/(2P^+)\), with outgoing helicity \(\lambda'\) and
incoming helicity \(\lambda\). The angular condition fixes \(H_{11}\) in terms of the three
independent amplitudes \(H_{00}\), \(H_{10}\) and \(H_{1,-1}\). With the helicity phases of
Ref.~\cite{KobushkinPQCD}, the form factors are
\begin{align}
 G_C&=\frac{2}{1+2\eta}\left(\frac{3-2\eta}{6}H_{00}
       +\frac{8}{3}\sqrt{\frac{\eta}{2}}H_{10}+\frac{2\eta-1}{3}H_{1,-1}\right),\notag\\
 G_M&=\frac{2}{1+2\eta}\left(H_{00}+\frac{2\eta-1}{\sqrt{2\eta}}H_{10}-H_{1,-1}\right),\notag\\
 G_Q&=\frac{2}{1+2\eta}\left(-\frac12H_{00}+\frac{H_{10}}{\sqrt{2\eta}}
       -\frac{1+\eta}{2\eta}H_{1,-1}\right).
 \label{eq:ff}
\end{align}
With the stated normalization Eq.~\eqref{eq:ff} gives \(G_C(0)=H_{00}(0)\) for a regular current.
The parametrizations below are used only in the fitted momentum range. At large \(Q\), the
helicity-conserving amplitude \(H_{00}\) is the leading one. The amplitudes \(H_{10}\) and
\(H_{1,-1}\) are suppressed by one and two additional powers of \(1/Q\),
respectively~\cite{CarlsonGross,KobushkinPQCD}.

\subsection{Parametrizations at finite momentum transfer}

We label \(00,10,1{-}1\) by \(i=0,1,2\) and define \(Q_\star^2=2\,\GeV^2\). The expression used for
the Born amplitudes is
\begin{equation}
 H_i(Q^2)=10^{-2}F_i(Q^2)
    \frac{G_D^2(Q^2/4)}{G_D^2(Q_\star^2/4)}
    \left(\frac{Q_\star}{Q}\right)^i\mathcal L_i(Q^2),
 \qquad
 G_D(Q_N^2)=\left(1+\frac{Q_N^2}{0.71\,\GeV^2}\right)^{-2}.
 \label{eq:heli}
\end{equation}
The common factor \(G_D^2\) represents the two nucleon dipoles at \(Q_N^2=Q^2/4\). The powers of
\(Q_\star/Q\) give the relative asymptotic suppression. The QCD factor \(\mathcal L_i\) is
normalized to one at \(Q_\star^2\). Hence \(F_i(Q_\star^2)=h_i\) gives
\(H_i(Q_\star^2)=10^{-2}h_i\).

For Table~\ref{tab:shapes} we use \(z=Q_\star^2/Q^2-1\), \(U_r(Q^2)=1+r(Q^2/Q_\star^2-1)\) and
\(L(Q^2)=\ln(Q^2/\Lambda_{\rm QCD}^2)\), with \(\Lambda_{\rm QCD}=0.2\,\GeV\) and
\(L_\star=L(Q_\star^2)\). The first form, \(f_1\), has one common power correction. In \(f_2\),
writing \(r_i=Q_\star^2/(Q_\star^2+M_i^2)\) makes \(U_{r_i}\) proportional to \(Q^2+M_i^2\). The
\(M_i\) are effective mass scales. In \(f_3\) the mass scale is common, and the fitted power \(k\)
shifts the logarithmic slope of all three amplitudes. At \(k=0\) this becomes \(f_2\) with equal
masses.

The term \(b_i z\) in \(f_4\) and \(f_5\) permits a different slope for each amplitude about
\(Q_\star^2\) at the same \(h_i\). In \(f_4\) the mass scale is common, and in \(f_5\) the three
masses are independent. Setting \(b_i=0\) in \(f_5\) recovers \(f_2\). For generic coefficients
\(F_i\) falls as \(1/Q^2\) up to logarithms. With the two dipoles, Eq.~\eqref{eq:heli} then gives
the expected \(Q^{-10-i}\) behavior. At finite \(Q^2\) the coefficients \(b_i\) account for
subleading terms phenomenologically.

\begin{table}[htbp]
\centering\small
\setlength{\tabcolsep}{4pt}
\caption{The five Born forms in Eq.~\eqref{eq:heli}. The count includes the three normalizations $h_i$.}
\label{tab:shapes}
\begin{tabular}{clp{.39\linewidth}r}
\toprule
Form & $F_i(Q^2)$ & Free shape terms & Parameters \\
\midrule
$f_1$ & $h_i(Q_\star^2/Q^2)(1+d z)$ & Common power correction & 4 \\
$f_2$ & $h_i/U_{r_i}$ & Separate mass scales & 6 \\
$f_3$ & $h_i(L/L_\star)^{-k}/U_{r_0}$ & Common mass and free logarithmic slope & 5 \\
$f_4$ & $(h_i+b_i z)/U_{r_0}$ & Independent $b_i$, common mass & 7 \\
$f_5$ & $(h_i+b_i z)/U_{r_i}$ & Independent $b_i$ and masses & 9 \\
\bottomrule
\end{tabular}
\end{table}

\subsection{One-loop anomalous dimensions and QCD logarithms}

We use the convention of Eq.~\eqref{eq:rge}, in which an eigencomponent of the distribution
amplitude evolves as \(\alpha_s^{\delta}\), with \(\delta=\gamma^{(0)}/\beta_0\). For \(n_f=3\)
active flavors, \(\beta_0=9\). The leading nucleon coefficient is \(\gamma_N^{(0)}=2/3\), so that
\(\delta_N=2/27\)~\cite{Braun}. Appendix~\ref{app:rg} gives the one-loop spectra of the local
isoscalar six-quark operators without derivatives. Their lowest eigenvalues are \(1\) at
\(\lambda=0\) and \(2/3\) at \(|\lambda|=1\). The corresponding coefficients of the Young-projected
evolution of Ji and Brodsky are \(1\) and \(7/6\)~\cite{JiPRD,JiReport}. The two calculations give
\((\delta_0^d,\delta_1^d)=(1/9,2/27)\) and \((1/9,7/54)\), respectively, where the subscript \(1\)
refers to \(|\lambda|=1\).

The leading hard six-quark amplitude carries \(\alpha_s^5\). Dividing by the two leading nucleon
form factors removes four powers from their hard amplitudes and \(4\delta_N\) from their
distribution amplitudes. At a common scale we obtain
\begin{align}
 \Gamma_{\lambda'\lambda}
   &=1+\delta^d_{\lambda'}+\delta^d_\lambda-4\delta_N,\notag\\
 (\Gamma_{00},\Gamma_{10},\Gamma_{1,-1})_{\rm local}
   &=\left(\frac{25}{27},\frac89,\frac{23}{27}\right),\notag\\
 (\Gamma_{00},\Gamma_{10},\Gamma_{1,-1})_{\rm JB}
   &=\left(\frac{25}{27},\frac{17}{18},\frac{26}{27}\right).
 \label{eq:powers}
\end{align}
Here \(\delta^d_{-1}=\delta^d_1\). For a single scale we take \(\mathcal
L_i=(L/L_\star)^{-\Gamma_i}\). If the nucleon factors are evaluated at \(Q^2/4\), we keep
\(L_4(Q^2)=\ln[Q^2/(4\Lambda_{\rm QCD}^2)]\) and use
\begin{equation}
 \mathcal L_i(Q^2)=\left(\frac{L}{L_\star}\right)^{-\Gamma_i}
   \left[\frac{L_4/L}{L_4(Q_\star^2)/L_\star}\right]^{4+4\delta_N}.
 \label{eq:twoscales}
\end{equation}
We also test the unreduced logarithm by restoring the factor \([L_4/L_4(Q_\star^2)]^{-4-4\delta_N}\)
while keeping the dipole envelope. For \(H_{00}\) the resulting inverse-logarithm power is
\(5+2\delta_0^d\), provided the lowest eigencomponent has a nonzero deuteron matrix element and hard
coefficient. For the two flip amplitudes Eq.~\eqref{eq:powers} accounts for external-state evolution
with the same hard-coupling counting. Their complete logarithms require the matching and evolution
of higher-twist operators. The additional power \(k\) in \(f_3\) is fitted independently of these
one-loop inputs.

\section{Interference of one- and two-photon exchange}
\label{sec:tpe}

\subsection{Scattering amplitude and observables}

We use the reduced scattering matrix \(\Tmat_h=\Tmat_h^{(0)}+\Tmat_h^{(2\gamma)}\) in the Breit
frame, where \(h=\pm1\) is the sign of the electron helicity~\cite{Kobushkin2011}. The superscripts
\(0\) and \(2\gamma\) refer to the Born and two-photon terms. The Born matrix follows from
\(G_C,G_M,G_Q\). Two-photon exchange shifts these to the generalized form factors \(\mathcal
G_j=G_j+\Delta G_j\) and introduces \(g_1,g_2,g_3\). The matrix is given in Eqs.~\eqref{eq:matrix}
and~\eqref{eq:generalFF}. Its spin labels are projections on the momentum-transfer axis in the Breit
frame. The light-front helicities of Eq.~\eqref{eq:ff} enter through the Born form factors.

For unpolarized incident particles the unnormalized recoil density matrix has the following Born and
interference parts,
\begin{align}
 \Rmat_0&=\tfrac12\sum_h\Tmat_h^{(0)}\Tmat_h^{(0)\dagger},\notag\\
 \Rmat_1&=\tfrac12\sum_h
  \left[\Tmat_h^{(2\gamma)}\Tmat_h^{(0)\dagger}
       +\Tmat_h^{(0)}\Tmat_h^{(2\gamma)\dagger}\right].
 \label{eq:density}
\end{align}
The cross section is proportional to \(D_0+D_1\), where \(D_j=\tr\Rmat_j\). For the numerator of
\(t_{2m}\) we take \(N_j^{(2m)}=\re\tr(\Rmat_j\tau_{2m})\) with the spin tensors \(\tau_{2m}\)
defined in Eq.~\eqref{eq:tensors}. Expanding the ratio to interference order gives
\begin{equation}
 t^{(0)}_{2m}=\frac{N_0^{(2m)}}{D_0},\qquad
 \Delta t_{2m}=\frac{N_1^{(2m)}-t^{(0)}_{2m}D_1}{D_0}.
 \label{eq:linear}
\end{equation}
The second term in the numerator of \(\Delta t_{2m}\) comes from the change in the unpolarized rate
used to normalize the polarization.

For an ultrarelativistic electron the virtual-photon polarization \(\epsilon\) and the electron
scattering angle \(\theta_B\) in the Breit frame satisfy
\begin{equation}
 \epsilon^{-1}=1+2(1+\eta)\tan^2\frac{\theta_{\rm lab}}2,\qquad
 y=\sqrt{\frac{1-\epsilon}{1+\epsilon}}=\sin\frac{\theta_B}{2},
 \qquad c_B=\sqrt{1-y^2}.
 \label{eq:angles}
\end{equation}
The real parts of the two-photon amplitudes are sufficient for the four observables considered here
at interference order. In the normalization of Eq.~\eqref{eq:v}, write them as
\begin{equation}
 \bm d=(\Delta v_C,\Delta v_Q,\Delta v_M,\bar g_1,g_2,g_3),
 \qquad \bar g_1=\sqrt{2\eta/3}\,g_1 .
 \label{eq:d}
\end{equation}
Here \(\Delta\bm v\) is obtained from \(\Delta G_j\) with the same weights as in Eq.~\eqref{eq:v}.
In Eq.~\eqref{eq:generalFF}, \(g_1\) enters the matrix elements between adjacent spin projections
and \(g_2+hy\,g_3\) those between \(+1\) and \(-1\). The \(g_i\) terms cancel from \(D_1\) but
survive in the tensor traces. We write the resulting reduced cross section for \(\theta_{\rm
lab}<180^\circ\) in terms of the generalized structure functions,
\begin{align}
 S(Q^2,\theta_{\rm lab})
 &=\mathcal A(Q^2,\theta_{\rm lab})
   +\mathcal B(Q^2,\theta_{\rm lab})\tan^2\frac{\theta_{\rm lab}}2,\notag\\
 \mathcal A&=A_{1\gamma}+\Delta A,\qquad
 \mathcal B=B_{1\gamma}+\Delta B,\notag\\
 \Delta A&=2\bm v\cdot\Delta\bm v,\qquad
 \Delta B=4(1+\eta)v_M\Delta v_M .
 \label{eq:ab_interference}
\end{align}
Thus the angular dependence of \(\mathcal A\) and \(\mathcal B\) beyond Born comes from the same
\(\Delta\bm v\) that enters the tensor calculation.

\subsection{Parametrizations of the two-photon amplitude}

In Ref.~\cite[Eqs.~(2) and (3)]{Shevchuk}, the \(Q^2\) dependence of the two-photon contribution was
written through factors of the form \(1+\alpha a[\ln(Q^2/\lambda_{\rm EM}^2)]^b\) multiplying the
Born structure functions. In Ref.~\cite[Eqs.~(7) and (8)]{Previous}, logarithmic terms were added to
\(A_{1\gamma}\) and \(B_{1\gamma}\). We retain the electromagnetic scale in these works. The
logarithm \(\ln(Q^2/\lambda_{\rm EM}^2)\) changes sign inside our interval. To permit real
noninteger powers throughout it, we use
\begin{equation}
 \ell(Q^2)=\ln\!\left(1+\frac{Q^2}{\lambda_{\rm EM}^2}\right),\qquad
 \lambda_{\rm EM}^2=1.75\,\GeV^2,\qquad
 \phi_j(Q^2)=\left[\frac{\ell(Q^2)}{\ell(Q_\star^2)}\right]^{\beta_j},
 \quad j=E,M .
 \label{eq:emlog}
\end{equation}
The functions \(\phi_j\) are one at \(Q_\star^2\) and have the same leading logarithmic behavior at
large \(Q^2\) as the earlier powers, up to constant normalization. The common factor \(\alpha y\),
where \(\alpha=e^2/(4\pi)\), implements the charge-odd interference. With the signed continuation
\(y\to-y\) and the helicity sum, the cross-section and tensor interference terms reverse
sign~\cite{Rekalo}.

\paragraph{Common relative factor.}
Take \(\Tmat_h^{(2\gamma)}=\delta\,\Tmat_h^{(0)}\), with \(\delta=\alpha y a_E\phi_E\). The strength
\(a_E\) and the power \(\beta_E\) are fitted. Then \(\Rmat_1=2\delta\Rmat_0\) and the fractional
change in the cross section is \(2\delta\). The correction in Eq.~\eqref{eq:linear} vanishes for
each tensor moment at fixed Born form factors. In a joint fit the tensor moments can still change
when the Born parameters are readjusted.

\paragraph{Separate electric and magnetic factors.}
Take the relative change to be \(\delta_E=\alpha y a_E\phi_E\) for the two electric form factors and
\(\delta_M=\alpha y a_M\phi_M\) for the magnetic form factor. The four-parameter choice is
\begin{equation}
 \bm d=\bigl(\delta_E v_C,\,\delta_E v_Q,\,\delta_M v_M,\,0,0,0\bigr).
 \label{eq:rel}
\end{equation}
The observable corrections in Eqs.~\eqref{eq:ab_interference} and~\eqref{eq:linear} now include a
change in the relative electric and magnetic weights. In particular,
\begin{equation}
 \Delta A=2\delta_E A_E+2\delta_M A_M,\qquad
 \Delta B=2\delta_M B_{1\gamma}.
 \label{eq:ab_relative}
\end{equation}
Equal strengths and powers give the common relative factor. We also fit the two strengths with the
powers fixed at \(\beta_E=\beta_M=0\) or \(1\). For zero powers only the angular factor remains in
the relative corrections.

\paragraph{Six spin amplitudes.}
To vary \(\Delta G_C,\Delta G_Q,\Delta G_M\) and \(g_1,g_2,g_3\) independently in strength we use
\begin{equation}
 \bm d=\alpha y\,\mathcal N(Q^2)
 \left(a_C\phi_E,\,a_Q\phi_E,\,a_M\phi_M,\,
       c_B^2a_1\phi_M,\,c_Ba_2\phi_E,\,c_Ba_3\phi_E\right).
 \label{eq:six}
\end{equation}
The factors of \(c_B\) give a regular backward limit for the extra spin terms. With six strengths
and two free powers there are eight parameters. We use \(U(Q^2)=\sqrt{A_{1\gamma}(Q^2)}\) and two
normalizations,
\begin{align}
 \mathcal N_{\rm local}(Q^2)&=U(Q^2),\notag\\
 \mathcal N_{\rm ref}(Q^2)&=U(Q_\star^2)\,
           \frac{G_D^2(Q^2/4)}{G_D^2(Q_\star^2/4)}\frac{Q_\star^2}{Q^2}.
 \label{eq:norms}
\end{align}
The local choice follows the Born strength at each \(Q^2\). The reference-point choice matches it at
\(Q_\star^2\) and has a fixed dipole envelope multiplied by \(Q_\star^2/Q^2\). In both cases an
additive \(\Delta v_M\) can remain finite at a zero of \(G_M\). The correction \(\Delta B\) still
vanishes there because it is proportional to \(G_M\Delta G_M\).

To compare the Born forms with the same two-photon dependence we use Eq.~\eqref{eq:six} with
\(\mathcal N=U\) and replace \(\phi_E,\phi_M\) by the common factor \((L/L_\star)^{-1}\). We also
examine a constant in place of this factor, and use the fixed powers \(\beta_E=\beta_M=1\) to
compare the two normalizations of Eq.~\eqref{eq:norms}.

\section{Experimental data and fit procedure}
\label{sec:data}

For \(Q^2\) between \(1\) and \(6\,\GeV^2\) we use 48 differential cross sections from the
world-data compilation, Jefferson Lab Halls~A and~C, and SLAC
NE4~\cite{AbbottPhen,Arnold,Martin,Auffret,Cramer,Alexa,Suleiman,AbbottA,Bosted,BostedReport}. We
include the Hall~C measurements of \(t_{20}\), \(t_{21}\) and \(t_{22}\) at four kinematic
settings~\cite{AbbottTensor}, for a total of 60 points. Lowering the cut to \(0.65\,\GeV^2\) gives
91 cross sections and eight measurements of each tensor moment. We fit the dimensional cross
sections at their laboratory angles and count each cross section once. The extracted structure
functions from the same measurements are therefore excluded from the fit. For NE4 the reported
acceptance determines an effective angle near \(175^\circ\) for each target length. Details of the
data selection and the cross-section units are given in Appendix~\ref{app:data}.

For \(N\) observations we use
\begin{equation}
 \chi^2=\sum_{\ell=1}^{N}
       \left(\frac{O_\ell^{\rm fit}-O_\ell^{\rm exp}}{\sigma_\ell}\right)^2,
 \qquad \nu=N-N_{\rm par}.
 \label{eq:chisq}
\end{equation}
Here \(O_\ell^{\rm exp}\) is a measured cross section or tensor moment, \(O_\ell^{\rm fit}\) its
fitted value and \(\sigma_\ell\) its quoted point uncertainty. We add the statistical and systematic
parts in quadrature where they are given separately. The parameter count \(N_{\rm par}\) includes
the Born and two-photon parameters. In the normalization check we let shared archival normalizations
vary with one Gaussian constraint per extra parameter, so the nominal \(\nu\) remains the same. The
tables show both \(\chi^2\) and \(\chi^2/\nu\). The experimental covariance is only partly known,
several fit parameters reach their bounds, and a two-photon power is undetermined when its strength
vanishes. We use \(\chi^2/\nu\) as a measure of the residuals per nominal degree of freedom.

To limit the size of the two-photon term at the measured kinematics we use the ratio
\begin{equation}
 \rho=\sqrt{D_2/D_0},\qquad
 D_2=\tfrac12\sum_h\tr\bigl(\Tmat_h^{(2\gamma)}\Tmat_h^{(2\gamma)\dagger}\bigr).
 \label{eq:rho}
\end{equation}
In the main fits we require \(\rho\le0.30\) and use \(|\alpha a_j|\le0.05\), \(|\beta_j|\le3\). The
omitted square contributes at most \(0.09D_0\) to the unpolarized trace at these settings. The
interference itself can be larger since \(|D_1|/D_0\le2\rho\). We vary these bounds to check how
they affect the fit.

\section{Comparison with experiment}
\label{sec:results}

\subsection{Dependence on the one-photon parametrization}

In Table~\ref{tab:bornfits} each Born form is fitted with a six-coefficient two-photon term and a
common inverse logarithm. For the reduced QCD logarithm at a single scale, the three independent
masses in \(f_2\) improve on the common shape in \(f_1\) and on the common mass with a free power in
\(f_3\). Giving each numerator its own \(b_i z\) term lowers the local-kernel \(\chi^2\) from
\(209.04\) for \(f_2\) to \(118.45\) for \(f_5\).

Taking the nucleon scale to be \(Q^2/4\) as in Eq.~\eqref{eq:twoscales} lowers the \(f_5\) value
further, to \(89.20\). The unreduced logarithm gives \(108.50\) with the same \(f_5\) form. For
\(f_2\), the separate scales increase \(\chi^2\). Their effect therefore depends on the
finite-\(Q^2\) terms. Once the \(f_5\) parameters are refitted, using the Ji--Brodsky powers in
place of the local powers changes \(\chi^2\) by less than \(0.10\) in the main separate-factor and
six-amplitude fits. The data in this interval constrain the momentum dependence of the Born
amplitudes much more strongly than they distinguish the two sets of anomalous dimensions.

\begin{table}[htbp]
\centering\small
\setlength{\tabcolsep}{4pt}
\caption{Fits of the Born forms to the same 60 measurements. Each fit uses six two-photon coefficients with a common inverse logarithm.}
\label{tab:bornfits}
\begin{tabular}{@{}llrrrrr@{}}
\toprule
 & & & \multicolumn{2}{c}{Local kernel} & \multicolumn{2}{c}{Ji--Brodsky} \\
\cmidrule(lr){4-5}\cmidrule(lr){6-7}
Form & Born logarithms & $\nu$ & $\chi^2$ & $\chi^2/\nu$ & $\chi^2$ & $\chi^2/\nu$ \\
\midrule
$f_1$ & Reduced, $L_4=L$ & 50 & 542.35 & 10.847 & 560.35 & 11.207 \\
$f_2$ & Reduced, $L_4=L$ & 48 & 209.04 & 4.355 & 211.56 & 4.408 \\
$f_3$ & Reduced, $L_4=L$ & 49 & 625.66 & 12.769 & 646.04 & 13.184 \\
$f_4$ & Reduced, $L_4=L$ & 47 & 141.79 & 3.017 & 141.84 & 3.018 \\
$f_5$ & Reduced, $L_4=L$ & 45 & 118.45 & 2.632 & 118.64 & 2.637 \\
\addlinespace
$f_2$ & Reduced with $L_4/L$ & 48 & 380.29 & 7.923 & 379.77 & 7.912 \\
$f_4$ & Reduced with $L_4/L$ & 47 & 108.55 & 2.310 & 109.20 & 2.323 \\
$f_5$ & Reduced with $L_4/L$ & 45 & 89.20 & 1.982 & 89.20 & 1.982 \\
\addlinespace
$f_2$ & Unreduced & 48 & 1413.72 & 29.452 & 1440.49 & 30.010 \\
$f_4$ & Unreduced & 47 & 152.56 & 3.246 & 160.86 & 3.423 \\
$f_5$ & Unreduced & 45 & 108.50 & 2.411 & 109.47 & 2.433 \\
\bottomrule
\end{tabular}
\end{table}

\subsection{Size of the two-photon contribution}

Table~\ref{tab:tpefits} compares two-photon parametrizations with the same \(f_5\) Born form, the
local spectrum and separate QCD scales. The one-photon fit gives \(\chi^2=93.14\) for 51 nominal
degrees of freedom. The common relative factor lowers it to \(87.71\) for 49. Its direct
interference changes only the cross section, as shown in Sec.~\ref{sec:tpe}. The change in the
tensor part of \(\chi^2\) in that row comes from the refitted Born amplitudes.

With separate electric and magnetic factors \(\chi^2\) is \(87.22\) for 47 nominal degrees of
freedom. The decrease of \(0.49\) relative to the common factor uses two additional parameters. The
ratio \(\chi^2/\nu\) changes from \(1.790\) to \(1.856\). In the separate-factor fit the
cross-section part decreases from \(84.07\) to \(80.69\) and the tensor part from \(9.07\) to
\(6.54\), compared with the one-photon fit. The six-amplitude fit with local normalization and two
free powers gives the smallest raw \(\chi^2\) in the table, \(86.09\), with \(\chi^2/\nu=2.002\).
With reference-point normalization the corresponding \(\chi^2\) is \(86.31\). In both six-amplitude
fits most of the decrease in \(\chi^2\) comes from the cross sections.

\begin{figure}[htbp]
\centering
\includegraphics[width=\linewidth]{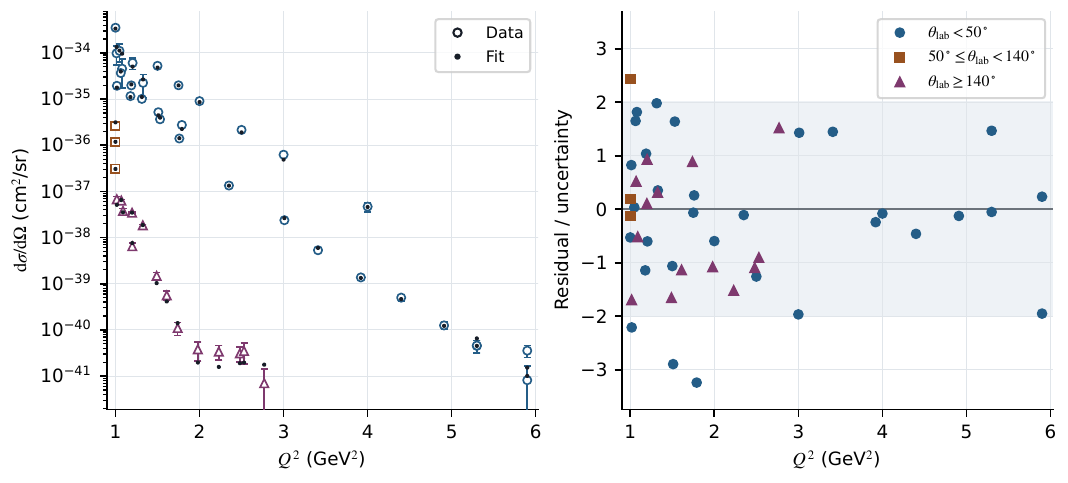}
\caption{Differential cross sections and residuals for \(f_5\) with separate electric and magnetic two-photon factors. Each point is evaluated at its laboratory angle, using the effective angles for NE4. The shaded band marks two point uncertainties on either side of zero.}
\label{fig:cross}
\end{figure}

\begin{figure}[htbp]
\centering
\includegraphics[width=\linewidth]{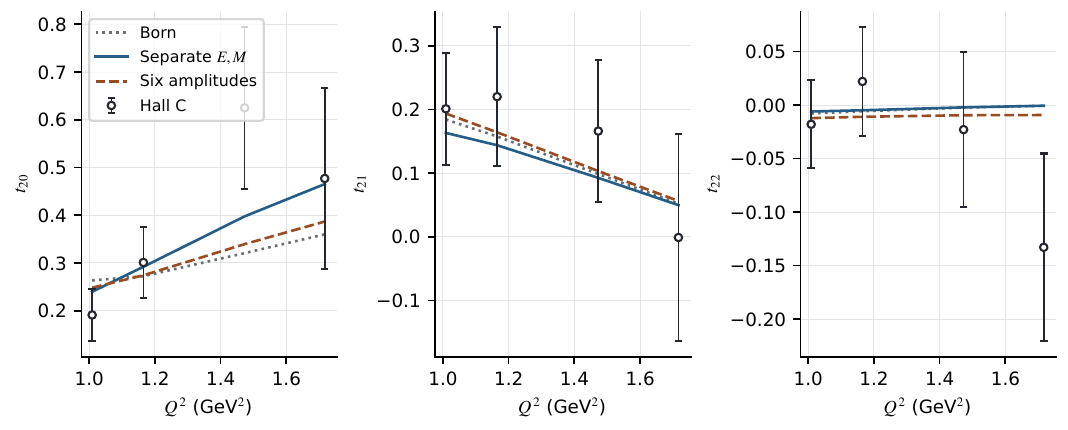}
\caption{Recoil tensor moments from Hall~C. The laboratory angles in order of increasing \(Q^2\) are \(29.8^\circ\), \(27.3^\circ\), \(23.0^\circ\) and \(19.8^\circ\). Lines join the values calculated at these settings for three separately fitted \(f_5\) parametrizations.}
\label{fig:tensor}
\end{figure}

Figures~\ref{fig:cross} and~\ref{fig:tensor} compare the fits with experiment at the actual
kinematic settings. The \(t_{22}\) numerator retains terms linear in \(g_1\) and \(g_2\), as
displayed in Eq.~\eqref{eq:numerators}. The present uncertainties in \(t_{22}\) still allow the
curves in Fig.~\ref{fig:tensor} to lie close to one another even when the cross-section corrections
differ.

\begin{table}[htbp]
\centering\small
\setlength{\tabcolsep}{4pt}
\caption{Fits of the same 60 observations with $f_5$ and separate QCD scales. $n_{2\gamma}$ counts the free two-photon coefficients and powers. $\chi^2_t$ combines the three tensor contributions.}
\label{tab:tpefits}
\begin{tabular}{@{}lrrrrrrr@{}}
\toprule
Interference form & $n_{2\gamma}$ & $\chi^2_\sigma$ & $\chi^2_t$ & $\chi^2$ & $\nu$ & $\chi^2/\nu$ &
$\max\rho$ \\
\midrule
Born & 0 & 84.07 & 9.07 & 93.14 & 51 & 1.826 & 0.000 \\
Common inverse log & 6 & 81.23 & 7.97 & 89.20 & 45 & 1.982 & 0.074 \\
Common constant & 6 & 81.09 & 8.08 & 89.17 & 45 & 1.982 & 0.082 \\
Common relative factor & 2 & 79.43 & 8.28 & 87.71 & 49 & 1.790 & 0.163 \\
Separate $E,M$, $\beta_E=\beta_M=0$ & 2 & 80.79 & 8.51 & 89.30 & 49 & 1.822 & 0.050 \\
Separate $E,M$, $\beta_E=\beta_M=1$ & 2 & 80.58 & 8.74 & 89.32 & 49 & 1.823 & 0.062 \\
Separate $E,M$, free powers & 4 & 80.69 & 6.54 & 87.22 & 47 & 1.856 & 0.158 \\
Local norm, $\beta_E=\beta_M=1$ & 6 & 80.25 & 8.24 & 88.49 & 45 & 1.966 & 0.116 \\
Local norm, free powers & 8 & 78.63 & 7.46 & 86.09 & 43 & 2.002 & 0.270 \\
Reference norm, $\beta_E=\beta_M=1$ & 6 & 79.74 & 7.98 & 87.72 & 45 & 1.949 & 0.121 \\
Reference norm, free powers & 8 & 78.96 & 7.35 & 86.31 & 43 & 2.007 & 0.300 \\
\bottomrule
\end{tabular}
\end{table}


\subsection{Dependence on the bounds and momentum range}

The separate-factor fit reaches both strength bounds and both power bounds, with \((\alpha
a_E,\alpha a_M)=(-0.05,+0.05)\) and \((\beta_E,\beta_M)=(3,-3)\). Reducing the power range to
\(|\beta_j|\le2\) raises \(\chi^2\) to \(88.72\). Allowing \(|\alpha a_j|\le0.10\) lowers it to
\(84.14\), with the largest \(\rho\) close to \(0.30\). This dependence on the imposed bounds limits
the accuracy with which the two powers can be inferred.

\begin{figure}[!htbp]
\centering
\includegraphics[width=\linewidth]{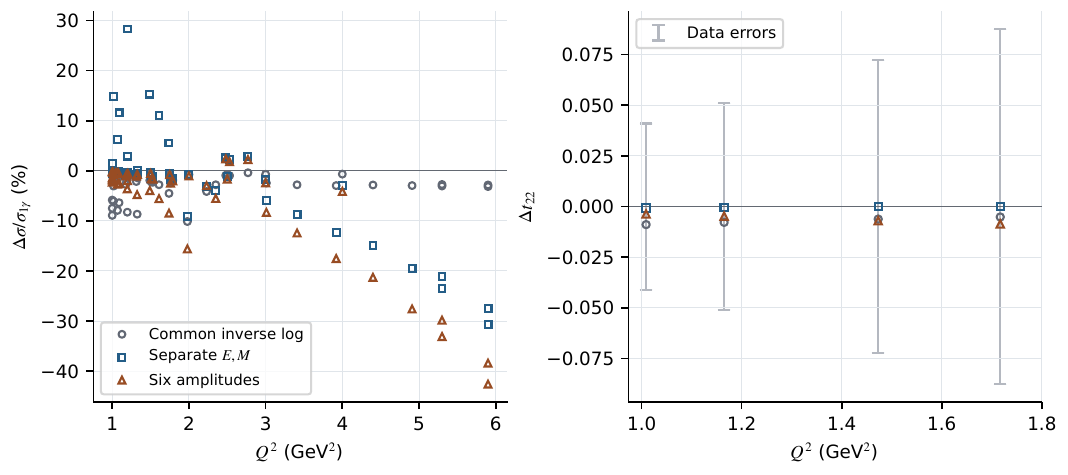}
\caption{Direct two-photon interference at fixed Born parameters. The left panel gives the fractional change in the differential cross section and the right panel gives the shift \(\Delta t_{22}\). The bars at zero in the right panel show the experimental uncertainties.}
\label{fig:fraction}
\end{figure}

For six amplitudes with local normalization, tightening the constraint to \(\rho\le0.15\) gives
\(\chi^2=87.14\). Table~\ref{tab:checks} also shows the effect of including the square of the real
model amplitude. The main fit scores change by less than \(0.3\). A full calculation at that order
would also include the imaginary parts of the two-photon amplitude and interference of one- and
three-photon exchange. Figure~\ref{fig:fraction} shows the direct shifts with the Born parameters
fixed at the values from each fit. The relative cross-section corrections in different fits can
differ by several tens of percentage points at the upper end of the range. The shifts in \(t_{22}\)
are small on the scale of the measured errors.

With the lower cut at \(0.65\,\GeV^2\), the common inverse-logarithm form gives \(\chi^2=282.91\)
for 100 nominal degrees of freedom. Six-amplitude fits with free powers give about \(224.81\) for
98. The tighter \(\rho\) bound gives approximately \(254.4\) for 98 with either normalization. With
local normalization, for example, \(\beta_E\) changes from \(+3\) to about \(-2.47\) when the lower
cut is moved. The change indicates sensitivity to the fitted momentum interval as well as to the
assumed bounds.

\begin{table}[htbp]
\centering\small
\setlength{\tabcolsep}{4pt}
\caption{Sensitivity of the $f_5$ fits. Each entry gives $\chi^2$ and $\nu$, with $\chi^2/\nu$ in parentheses below.}
\label{tab:checks}
\begin{tabular}{@{}clrrr@{}}
\toprule
$Q^2_{\rm min}\,(\GeV^2)$ & Check & Separate $E,M$ & Local norm & Reference norm \\
\midrule
1.00 & Primary bounds & \shortstack[r]{87.22, 47\\(1.856)} & \shortstack[r]{86.09, 43\\(2.002)} & \shortstack[r]{86.31, 43\\(2.007)} \\
1.00 & $|\beta_j|\le2$ & \shortstack[r]{88.72, 47\\(1.888)} & --- & --- \\
1.00 & $|\alpha a_j|\le0.10$ & \shortstack[r]{84.14, 47\\(1.790)} & \shortstack[r]{84.48, 43\\(1.965)} & \shortstack[r]{85.25, 43\\(1.983)} \\
1.00 & $\rho\le0.15$ & --- & \shortstack[r]{87.14, 43\\(2.027)} & \shortstack[r]{87.11, 43\\(2.026)} \\
1.00 & Archival normalization & \shortstack[r]{86.77, 47\\(1.846)} & --- & --- \\
1.00 & Square of model amplitude & \shortstack[r]{87.27, 47\\(1.857)} & \shortstack[r]{86.01, 43\\(2.000)} & \shortstack[r]{86.05, 43\\(2.001)} \\
0.65 & Primary bounds & \shortstack[r]{263.25, 102\\(2.581)} & \shortstack[r]{224.81, 98\\(2.294)} & \shortstack[r]{224.81, 98\\(2.294)} \\
0.65 & $\rho\le0.15$ & --- & \shortstack[r]{254.37, 98\\(2.596)} & \shortstack[r]{254.36, 98\\(2.596)} \\
\bottomrule
\end{tabular}
\end{table}

\FloatBarrier

\subsection{Relation to our earlier fit}

In Ref.~\cite{Previous} the \(f_1\) fit gave \(\chi^2=18.40\) with 16 nominal degrees of freedom,
and the Hall~C tensor triplet was shown at \(Q^2=1.717\,\GeV^2\). All fits in
Tables~\ref{tab:bornfits} and~\ref{tab:tpefits} use a common sample of differential cross sections
at their measured angles and all four Hall~C tensor triplets. On these data \(f_5\) describes the
measurements better than \(f_2\) and \(f_4\) with each of the three free-power interference models
at separate QCD scales. The improvement establishes the importance of the additional finite-\(Q^2\)
freedom within this set of parametrizations. Its microscopic origin would require an independent
calculation of subleading contributions.

\section{Conclusions}

The one-loop kernel for local isoscalar six-quark operators without derivatives gives the spectra
\(\{1,4\}\) and \(\{2/3,8/3\}\) for \(\lambda=0\) and \(|\lambda|=1\), respectively. The lowest
eigenvalues fix the external-state contribution to the logarithmic powers. The local and Ji--Brodsky
inputs lead to nearly the same fit after the Born amplitudes are readjusted. Independent
finite-\(Q^2\) terms in those amplitudes give a substantially larger improvement in the description
of cross sections and recoil polarization.

We have evaluated the two-photon interference in the cross section and the three tensor observables
using the same Breit-frame amplitude. A common relative factor accounts for most of the decrease in
\(\chi^2\) obtained with separate electric and magnetic factors. The small further decrease, the
boundary solutions and the changes with the momentum cut leave the separate two-photon logarithmic
powers poorly determined.

\paragraph{Data and code availability.}
The measurements are from the cited publications. The supplementary files provide the data
selection, fit results and code used to calculate the local spectrum and reproduce the tables and
figures.

\paragraph{Use of generative AI.}
The authors used ChatGPT (OpenAI) for language editing, translation and assistance with numerical
implementation and verification. The authors take responsibility for the manuscript.

\appendix

\section{One-loop evolution of the local operators}
\label{app:rg}

For a distribution-amplitude eigencomponent $\Phi_a$ we use the convention
\begin{align}
 \mu\frac{\dd\Phi_a}{\dd\mu}&=-\frac{\as}{2\pi}\gamma_a^{(0)}\Phi_a,&
 \mu\frac{\dd\as}{\dd\mu}&=-\frac{\beta_0}{2\pi}\as^2+O(\as^3),&
 \beta_0&=11-\tfrac23 n_f .
 \label{eq:rge}
\end{align}

Dividing the two evolution equations gives \(\dd\ln\Phi_a/\dd\ln\alpha_s=\gamma_a^{(0)}/\beta_0\).
Thus \(\Phi_a\) has the power \(\delta_a=\gamma_a^{(0)}/\beta_0\) of \(\alpha_s\). The one-loop
kernel eigenvalues below use \(\alpha_s/(2\pi)\) as in Eq.~\eqref{eq:rge}. A convention using
\(\alpha_s/(4\pi)\) assigns coefficients twice as large and gives the same \(\delta_a\). We use
\(N_c=3\) and \(C_F=4/3\), and take the \(u,d\) quarks in the operators to be massless.

Let \(q^{af}_\zeta\) be a good light-front quark component of helicity \(\zeta=\pm1/2\), color \(a\)
and flavor \(f\). At the lowest local moment the fields coincide and there are no derivatives. Each
field has collinear conformal spin \(j=1\), so the two-quark conformal spin is \(J_{ij}=2\). Using
\(\psi(2)-\psi(1)=1\), \(\psi(3)-\psi(1)=3/2\) and the quark constant \(\sigma_q=3/4\), the pair
kernels of Ref.~\cite[Eqs.~(3.3), (4.23) and (4.24)]{BMR} reduce to
\begin{align}
 \mathbb H^{\rm same}_{ij}
 &=-4\,\bm T_i\cdot\bm T_j\,[\psi(2)-\psi(1)-3/4]
  =-\bm T_i\cdot\bm T_j ,\notag\\
 \mathbb H^{\rm opp}_{ij}
 &=-2\,\bm T_i\cdot\bm T_j\,[\psi(3)-\psi(1)-3/2]=0 .
 \label{eq:pairs}
\end{align}

Here \(\psi\) is the digamma function and \(\bm T_i\) is a color generator acting on quark \(i\),
with \(\tr(T^aT^b)=\delta^{ab}/2\) in the fundamental representation. The opposite-helicity pair
consists of the two quark components of the Dirac field, \(\psi_+\) and \(\bar\chi_+\) in the
notation of Ref.~\cite{BMR}. The additional annihilation terms of its Eq.~(4.25) belong to the
quark--antiquark channels. The field-renormalization terms are already included in
Eq.~\eqref{eq:pairs}. Put \(\bm T_\pm=\sum_{i:\zeta_i=\pm1/2}\bm T_i\) for the two helicity groups.
The sum over equal-helicity pairs follows by squaring the two group generators. For \(n\) quarks
this gives
\begin{equation}
 \mathbb H=\frac{n C_F-\bm T_+^{\,2}-\bm T_-^{\,2}}{2}
           =\frac{n C_F}{2}-\bm T_+^{\,2}\quad\text{on color singlets}.
 \label{eq:casimir}
\end{equation}

The equality on the color-singlet subspace follows from \((\bm T_++\bm T_-)|1\rangle=0\), where
\(|1\rangle\) denotes a color-singlet state. With three quarks and total helicity \(1/2\), the
equal-helicity pair must have color \(\bar{\mathbf3}\) so that it combines with the third quark to a
singlet. Its Casimir is \(C_F\), giving \(\gamma_N^{(0)}=3C_F/2-C_F=2/3\), consistent with the
leading nucleon eigenvalue of Ref.~\cite{Braun}.

For \(k\) same-helicity quarks the color and flavor indices of the local operator are antisymmetric.
They therefore lie in \(\wedge^k(\mathbf3,\mathbf2)\) under \(SU(3)_{\rm c}\times SU(2)_I\). The
needed decompositions are
\begin{align}
 \wedge^2(\mathbf3,\mathbf2)&=(\bar{\mathbf3},\mathbf3)\oplus(\mathbf6,\mathbf1),\notag\\
 \wedge^3(\mathbf3,\mathbf2)&=(\mathbf1,\mathbf4)\oplus(\mathbf8,\mathbf2),\notag\\
 \wedge^4(\mathbf3,\mathbf2)&=(\mathbf3,\mathbf3)\oplus(\bar{\mathbf6},\mathbf1).
 \label{eq:exterior}
\end{align}
The first entry of each pair denotes color and the second the dimension of the isospin multiplet. A
color singlet requires conjugate colors in the two helicity groups, and $I=0$ requires equal
isospins. The local operators have quark helicity sum $\lambda=(n_+-n_-)/2$. The two allowed color
pairings for each $|\lambda|$ and the eigenvalue $4-C_2(R_+)$ of each pairing are shown in
Table~\ref{tab:operators}.

\begin{table}[htbp]
\centering
\caption{One-loop spectrum of the lowest local, isoscalar six-quark color singlets. $C_2(R_+)$ is the quadratic Casimir of the positive-helicity group.}
\label{tab:operators}
\begin{tabular}{ccccc}
\toprule
$|\lambda|$ & $(n_+,n_-)$ & $R_+\otimes R_-$ & $C_2(R_+)$ & $\gamma^{(0)}$ \\
\midrule
$0$ & $(3,3)$ & $\mathbf1\otimes\mathbf1$ & $0$ & $4$ \\
    &         & $\mathbf8\otimes\mathbf8$ & $3$ & $1$ \\
$1$ & $(4,2)$ & $\mathbf3\otimes\bar{\mathbf3}$ & $4/3$ & $8/3$ \\
    &         & $\bar{\mathbf6}\otimes\mathbf6$ & $10/3$ & $2/3$ \\
\bottomrule
\end{tabular}
\end{table}

Direct diagonalization gives the same spectra. We use a fermion occupation basis with three \(u\)
and three \(d\) quarks and restrict it to the common null space of the total color Casimir and the
isospin-raising operator.

To relate these results to Ref.~\cite{JiReport}, the sum of \(\bm T_i\cdot\bm T_j\) over the 15
pairs in a six-quark color singlet is \(-3C_F\). Replacing the opposite-helicity pair factors by the
average \(-C_F/5\) in \(\mathbb H=3C_F+\sum_{\rm opp}\bm T_i\cdot\bm T_j\) gives \(6C_F/5\) and
\(7C_F/5\), with nine and eight opposite-helicity pairs, respectively. These are the
symmetric-kernel values of Eq.~(5.16) of Ref.~\cite{JiReport} in our normalization. Its subsequent
Young projection gives \(1\) and \(7/6\) by Eq.~(5.17). The difference at unit helicity is between
that projected coefficient and the local spectrum of Table~\ref{tab:operators}. The fits use the two
sets as alternative inputs for the evolution of the external states.

The two external distribution amplitudes contribute $L^{-\delta^d_{\lambda'}-\delta^d_\lambda}$. The
leading hard scattering of six quarks carries five powers of
$\alpha_s$~\cite{BrodskyLepage,BrodskyJi,JiPRD}. At the same accuracy each nucleon form factor at
$Q^2/4$ is proportional to $(Q^2/4)^{-2}L_4^{-2-2\delta_N}$. Dividing by two such form factors gives
the logarithmic dependence

\begin{equation}
 L^{-5-\delta^d_{\lambda'}-\delta^d_\lambda} L_4^{4+4\delta_N}
 =L^{-\Gamma_{\lambda'\lambda}}(L_4/L)^{4+4\delta_N}.
\end{equation}

\section{Spin matrix and tensor traces}
\label{app:spin}

In the Breit frame the $z$ axis points along the transferred momentum. The $y$ axis points along
$\bm k\times\bm k'$ for the incoming and outgoing electron momenta. In the spin basis $(+1,0,-1)$
the common matrix $\Tmat_h=\Tmat_h^{(0)}+\Tmat_h^{(2\gamma)}$ reads~\cite{Kobushkin2011}
\begin{equation}
 \Tmat_h=
 \begin{pmatrix}
 c_B(\mathcal G_C-\frac23\eta\mathcal G_Q)&-X_h&Y_h\\
 X_{-h}&c_B(\mathcal G_C+\frac43\eta\mathcal G_Q)&-X_h\\
 Y_{-h}&X_{-h}&c_B(\mathcal G_C-\frac23\eta\mathcal G_Q)
 \end{pmatrix},
 \label{eq:matrix}
\end{equation}
\begin{align}
 X_h&=\sqrt{\eta/2}\,[(1+h y)\mathcal G_M+(y^2-hy)g_1],
 &Y_h&=g_2+h y g_3,&\mathcal G_j&=G_j+\Delta G_j\quad(j=C,Q,M).
 \label{eq:generalFF}
\end{align}

The unnormalized density of Eq.~\eqref{eq:density} gives the dimensional cross section

\begin{equation}
 \begin{aligned}
 \frac{\dd\sigma}{\dd\Omega}&=\frac{K}{3}(D_0+D_1),\qquad
 K=\frac{\alpha^2}{4E^2}\frac{E'}{E}
       \frac{1+\eta\sin^2(\theta_{\rm lab}/2)}{\sin^4(\theta_{\rm lab}/2)},\\
 E'&=E-\frac{Q^2}{2M_d},\quad Q^2=4EE'\sin^2(\theta_{\rm lab}/2).
\end{aligned}
 \label{eq:cross}
\end{equation}

Here $E$ and $E'$ are the electron energies in the laboratory frame. For $\theta_{\rm
lab}<180^\circ$, $Kc_B^2=\sigma_{\rm Mott}$ including recoil. The expression in terms of $K$ is
finite at $180^\circ$. We convert units using $1\,\GeV^{-2}=3.89379\times10^{-28}\,\mathrm{cm}^2$.
The tensor weights in Eq.~\eqref{eq:linear} are
\begin{equation}
 \tau_{20}=(3S_z^2-2I)/\sqrt2,\qquad
 \tau_{21}=-(\sqrt3/2)(S_+S_z+S_zS_+),\qquad
 \tau_{22}=(\sqrt3/2)S_+^2 .
 \label{eq:tensors}
\end{equation}

Here $I$ is the identity and $S_z$ and $S_+=S_x+\mathrm i S_y$ are spin-one generators.
Let $[XY]_{01}=\re(\mathcal G_X^*\mathcal G_Y)$ with terms through first order in $\Delta G_j$.
Products involving $g_i$ need only the Born form factors at this order. The spin traces in
Eq.~\eqref{eq:linear} give the following numerators $N_0+N_1$,
\begin{align}
 N_{20}^{(01)}={}&-\frac{\eta}{\sqrt2}
  \left(8c_B^2[CQ]_{01}+\frac83\eta c_B^2[QQ]_{01}+(1+y^2)[MM]_{01}\right),\notag\\
 N_{21}^{(01)}={}&\sqrt{3\eta}\left[2c_B\eta\big([MQ]_{01}+y^2G_Q\re g_1\big)
      +G_M\re(g_2+y^2g_3)\right],\notag\\
 N_{22}^{(01)}={}&\frac{\sqrt3}{2}\left[-\eta c_B^2[MM]_{01}-4\eta y^2G_M\re g_1
      +4c_B(G_C-\tfrac23\eta G_Q)\re g_2\right].
 \label{eq:numerators}
\end{align}
The corresponding unpolarized trace is
\begin{equation}
 D_0+D_1
 =3c_B^2\bigl[U^2+2\bm v\cdot\Delta\bm v\bigr]
  +6y^2\bigl[v_M^2+2v_M\Delta v_M\bigr].
 \label{eq:denominator}
\end{equation}
For \(\theta_{\rm lab}<180^\circ\), \(c_B^2\tan^2(\theta_{\rm lab}/2)=y^2/(1+\eta)\) brings this
expression to Eq.~\eqref{eq:ab_interference} in Mott units. At \(180^\circ\) the trace in
Eq.~\eqref{eq:denominator} is regular.

Expansion of the ratio gives Eq.~\eqref{eq:linear}. The target tensor analyzing powers use
$\Tmat_h^\dagger\Tmat_h$ in the spin trace. The real $21$ component changes sign between the target
and recoil conventions used here. The $20$ and $22$ components agree.

\section{Data treatment and numerical settings}
\label{app:data}

For Hall~A we use both cross-section tables in the E91-026 analysis, including the angles near
$144.5^\circ$~\cite{Alexa,Suleiman}. For Hall~C we use the cross sections of Table~I of
Ref.~\cite{AbbottA}. Reconstructing the cross sections from the tabulated $A$, $B$ and the Mott
factor gives units of $\mathrm{pb/sr}$ for the numerical entries, whereas the table header gives
$\mathrm{nb/sr}$. We use the relative errors quoted for $A$ as the available error estimate for
these cross sections. The Hall~C tensor values and angles are from Table~I of
Ref.~\cite{AbbottTensor}. This includes $t_{22}=+0.087$ at $Q^2=0.651\,\GeV^2$ in the extended fit.
We retain the CEA compilation's rejection flags and count the overlapping forward Arnold--Martin
kinematics once. The extension includes the archived Gar\c{c}on recoil data~\cite{Garcon}.

For NE4 we use the measured cross sections in Table~III of Ref.~\cite{BostedReport}, prior to
subtraction of the forward-structure contribution. For each target length we match the reported
acceptance average $\langle\cos^2(\theta/2)/\sin^4(\theta/2)\rangle$ to obtain an effective angle.
The angles are $175.0^\circ$, $175.1^\circ$ and $175.6^\circ$. This approximation models the leading
acceptance effect for these nearly backward measurements.

We fix the overall Born sign by $h_0\ge0$ and use $|h_i|,|d|\le5$, $|k|\le6$, $|b_i|\le3$ and
$0.005\le r_i\le1$. The bounds on the interference coefficients refer to $\alpha a_j$, with
$\alpha=e^2/(4\pi)$. Differential evolution and local least squares or constrained minimization are
used with starts from the Born and interference fits. The relative factors provide an independent
check of the constraint. At fixed kinematics the electric and magnetic Born spin matrices are
orthogonal under the spin trace, so $\rho$ cannot exceed $|y|\max_j |\alpha a_j\phi_j|$. On the 60
measured settings this upper bound is below $0.226$ for the main relative fit domain. The same
solution can therefore be checked by bounded least squares without applying the $0.30$ condition in
the optimizer. Several Born mass parameters and interference coefficients reach their imposed
bounds. The fitted parameter vectors and the residual at each measured angle are given in the
accompanying files.

\end{document}